\documentclass[letterpaper]{article}
\usepackage{aaai}
\usepackage{times}
\usepackage{helvet}
\usepackage{courier}
\usepackage[most]{tcolorbox}
\usepackage{color, colortbl}
\definecolor{Gray}{gray}{0.9}
\definecolor{LightGray}{gray}{0.6}
\newcommand{\laura}[1]{\textit{Laura:#1]}}

\begin{document}

\title{The Theory, Practice, and Ethical Challenges of \\ Designing a Diversity-Aware Platform for Social Relations}

\author{Laura Schelenz\\
International Center for Ethics \\in the Sciences and \\ Humanities (IZEW) \& \\ University of Tuebingen\\ Tuebingen, Germany\\
laura.schelenz@uni-tuebingen.de \And Ivano Bison\\
University of Trento \\Trento, Italy\\
ivano.bison@unitn.it \And
Matteo Busso\\
University of Trento\\ Trento, Italy\\
matteo.busso@unitn.it \And Amalia de Götzen\\
Aalborg University \\Copenhagen, Denmark\\
ago@create.aau.dkl \AND
Daniel Gatica-Perez\\
Idiap Research Institute \&\\École Polytechnique \\Fédérale de Lausanne (EPFL)\\ Switzerland\\
gatica@idiap.ch \And
Fausto Giunchiglia\\
University of Trento \\Trento, Italy\\
fausto.giunchiglia@unitn.it \And
Lakmal Meegahapola\\
Idiap Research Institute \&\\École Polytechnique \\Fédérale de Lausanne (EPFL)\\ Switzerland\\
lmeegahapola@idiap.ch \And Salvador Ruiz-Correa\\
Instituto Potosino de Investigacion \\ Cientifica y Tecnológica (IPICYT) \\San Luis Potosí, Mexico\\
salvador.ruiz@ipicyt.edu.mx}

\maketitle
\begin{abstract}
\begin{quote}
Diversity-aware platform design is a paradigm that responds to the ethical challenges of existing social media platforms. Available platforms have been criticized for minimizing users' autonomy, marginalizing minorities, and exploiting users' data for profit maximization. This paper presents a design solution that centers the well-being of users. It presents the theory and practice of designing a \textit{diversity-aware} platform for social relations. In this approach, the diversity of users is leveraged in a way that allows like-minded individuals to pursue similar interests or diverse individuals to complement each other in a complex activity. The end users of the envisioned platform are students, who participate in the design process. Diversity-aware platform design involves numerous steps, of which two are highlighted in this paper: 1) defining a framework and operationalizing the "diversity" of students, 2) collecting "diversity" data to build diversity-aware algorithms. The paper further reflects on the ethical challenges encountered during the design of a diversity-aware platform. 
\end{quote}
\end{abstract}

\section{1. Introduction}

Existing social media platforms exhibit a series of challenges for users' autonomy and well-being. Feedback loops and filter bubbles tend to curtail access to the breadth of information \cite{Pariser.2012}. Popular content and mainstream norms are picked up and further promoted by algorithms, effectively marginalizing alternative voices \cite{Brough.2020,Trillo.2018}. Extensive and often sensitive data is collected with little awareness for the conditions of informed consent \cite{Tufekci.19March2018}. This data is used for personalized advertisement by exploiting users' vulnerabilities \cite{Levin.1May2017}, which can amount to "manipulation" and impede users' free decision-making \cite{Susser.2019}. 

This paper presents on-going research into the design and implementation of a diversity-aware platform for social relations. In our project, we develop the culture, science, and engineering methodologies, algorithms, as well as social interaction protocols for said online platform. The WeNet platform 
seeks to empower machine mediated diversity-aware interactions between people. The technology builds on the diversity of students, who are the envisioned end users, to accommodate their diverse needs and preferences. It is designed around community help and volunteering, connecting people while also protecting their individual privacy, in accordance with the GDPR \cite{EuropeanParliament.2016}. 

Designing a diversity-aware platform involves several steps: the theoretical and ethical framework, the collection, processing, and preparation of data, the development and auditing of diversity-aware algorithms, and the development of the research and innovation infrastructure. Of these tasks, we highlight the first two in this paper 1) the definition and operationalization of “diversity,” and 2) the collection of “diversity” data. The diversity-aware design paradigm that we propose in this paper is novel. To our knowledge, there is currently no other platform that follows an explicitly diversity-oriented path. 

In the design of the WeNet platform and app, and for the sake of this paper, we define diversity as differences between users with regard to their social practices (routine behavior) and diversity-awareness as the skill to navigate diverse environments. The proposed platform is twice diversity-aware in the sense that it leverages the diverse practices of students to improve interaction in their community and helps students navigate diversity in their community via computational tools. In practice, this means implementing diversity-awareness in every platform component including the algorithms powering the coordination of the community, matching of users, recommendations, and incentives. 
Despite promises of this approach, engaging with diversity requires ethical framing and continuous reflection. Diversity classification are powerful tools that structure and shape communities and societies \cite{Ahmed.2012,Haslanger.2012}. In the discussion section of the paper, we therefore make ethical challenges visible to inform the on-going design process.

The diversity-aware platform design offered in this paper is situated in a European context. Although the project cooperates with universities around the world, it is primarily bound by European values and law. As such, the project is one of the first research endeavors that incorporates strict data protection (GDPR, \cite{EuropeanParliament.2016}) from the beginning of the design process. Furthermore, the public funding ensures a framework that centers public interest and the well-being of populations rather than commercial logic. Despite this framework, we recognize that ethical challenges remain and, again, attend to them in the discussion section.  

The paper is structured as follows: Section 2 summarizes challenges of existing social media platforms, highlighting how they suppress diversity and rely on problematic data collection processes. Section 3 presents our approach, whereas section 3.1 introduces the way that users' diversity is theorized and operationalized and section 3.2 describes the collection of diversity data from numerous pilot sites. Section 4 assesses the ethical challenges of diversity-aware platform design. The paper makes three  contributions: 1) recognizing the need for improved designs of social media platforms, 2) introducing a diversity-aware paradigm for ethical platform design, and 3) providing reflections on the ethical challenges of the new approach. 

\section{2. Problems with Existing \\ Social Media Platforms}

The Netflix documentary “The Social Dilemma” alarmingly points to systemic challenges in the design and use of social media \cite{Orlowski.2020}. Indeed, social media platforms like Facebook, Twitter, and YouTube have been heavily criticized for creating filter bubbles \cite{Pariser.2012}; the manipulation of users for economic benefits \cite{Zuboff.2019,Susser.2019,Bodle.2017}, the exploitation of users’ data \cite{Tufekci.19March2018,TatlowGolden.2020}, the distribution of hate speech and fake news \cite{Daniels.2009,Lumsden.2019}, and the potential for addiction \cite{Lembke.2020}. As a consequence of these practices, democratic principles such as information diversity have been said to be violated \cite{Marantz.2019}. Whereas this conglomeration of challenges exists, we focus on two aspects: the suppression of diversity in social media platforms and problematic data collection practices. 

There are numerous ways in which diversity is reduced in social media platforms: First, a well-recognized problem is the exposure of users to an unhealthy feedback loop, also known as the filter bubble (Pariser 2012).\footnote{Note that we refer to a lack of diversity \textit{within} a social media platform. Research that investigates the link of a filter bubble or echo chamber to segregation and polarization in society stresses that most users are still exposed to a variety of media outlets beyond social media \cite{Dubois.2018,Flaxman.2016}.} It encourages users to interact with content and users that are like-minded. This is in large part due to the operation of recommender algorithms, which connect users to users or users to content. When a user is new to a platform, there is little information as to what content they might like to consume. Since the assumption is that connected users ("friends") are like-minded, the recommender algorithm will show the new users what their connections see. A new user may then be forced to become similar to another user in the platform, which manifests homogeneity in an online community \cite{Sacharidis.2020}. 

Another problem is that popular content in platforms is picked up by algorithms and recommended to users, also called "popularity bias" \cite{Abdollahpouri.2019}. This reduces diversity of content, and becomes especially concerning if the content produced by particular social groups is constantly under-ranked \cite{Beutel.2019}. In the context of political exchange, the reduction of content diversity  can threaten a pluralism of opinion. In a democratic society, a lack of information diversity may stifle political discourse and alter political processes \cite{Bay.2018}. 

The lack of diversity in a platform has implications for users. Brough, Literat, and Ikin \shortcite{Brough.2020} found that youth felt the need to conform to standard community norms to be seen and recognized in social media platforms. Being different or posting about less covered topics is not rewarded by the system. Interviewed youth report: "I have a couple of other Latino friends posting about certain holidays or things like that. You don’t get the same reactions you would from posting things
about Christmas or Thanksgiving" \cite[p. 5]{Brough.2020}. 

Another concern for users' well-being is the proliferation of online violence in the form of hate speech, stalking, bullying, doxing, and revenge pornography ~\cite{Lumsden.2019,Segrave.2017}. Particularly minority groups may be targeted with harmful, racist content ~\cite{Daniels.2009}. This may discourage users from staying active in online communities. Unfortunately, social media platforms tend to reinforce and even exacerbate harmful social dynamics due to a range of reasons: lack of regulation on behalf of governments, lack of appropriate content moderation on behalf of social media sites, and weakness of law enforcement in the persecution of crimes online ~\cite{Lumsden.2019}. 

The second aspect we highlight in this section is data collection. Zuboff (2019) has voiced concern about the emergence of “surveillance capitalism” and refers to the practices and frameworks of social media platforms. Because services like Facebook and Google remain free, companies rely on advertisement to generate revenue. This advertisement is tied to users’ engagement: the longer users engage in a platform, the more ads users will see and the more revenue the company will gain. Platforms therefore use psychology and steer users' behavior to keep them engaged in the service. This practice is done by exploiting (sensitive) information about users by way of data collection. For instance, Facebook was found to target vulnerabilities of teenagers to maximize profit \cite{Levin.1May2017}.


This psychological targeting of users has implications for users' well-being and users' autonomy. 
Susser et al. (2019) warn of the consequences of “online manipulation” for the ability of a user to make an informed decision. By subtly changing designs and information based on the personal profile of a user (i.e. changing the “choice architecture,” p. 23), the user is affected in their decision-making power. Limiting the array of possibilities in front of users may limit their ability to envision different options. This has implications for users' autonomy and the ability of a citizen to determine their life plan, which is of fundamental value in a liberal democracy \cite[p. 35]{Susser.2019}. 


How can we address the challenges of existing social media platforms? On the one hand, there is a need for regulation and possibly the certification of social media platforms \cite{Heesen.2020}. Recent years saw an increase in regulation attempts in the European context \cite{EuropeanParliament.2016}. This move towards increased regulation may be read against the European Union’s intention of developing more “digital sovereignty” \cite{EuropeanCommission.19February2020}. European stakeholders also develop new approaches to operationalizing ethical principles \cite{AIEthicsImpactGroup.2020}. 

On the other hand, addressing the problems of existing social media platforms requires a radical rethinking of the values that inform the design of social media platforms. The challenges described in this section have been provoked by \textit{deliberate} design choices that seek to maximize user attention to the platforms. Social media platforms are not inherently or exclusively bad for users’ autonomy. On the contrary, social media platforms have been experienced as empowering spaces that allow individuals to express themselves and organize for political resistance ~\cite{Brough.2020,Williams.2015}. Hence, in our view, technology-mediated social interactions require new legal frameworks, new business models, and, most tangibly, new \textit{designs} in order to avoid the stated problems.

\section{3. A Diversity-aware Platform}

This section introduces our approach to designing a diversity-aware platform for social relations. The platform constitutes the foundational location for the development of diversity-aware applications that seek to improve student community life. In the following, we refer to the platform as WeNet platform. We also refer to the WeNet app, which is the primary research-driven prototype on top of the platform that will be tested in numerous pilot sites across the world. 

The design process for the WeNet paradigm can be characterized by 1) its value-sensitive approach and 2) its participatory approach. 
First, designers acknowledge that values are embedded consciously or unconsciously in technologies ~\cite{Friedman.1996}. According to Brey ~\shortcite{Brey.2010}, "the embedded values approach holds that computer systems and software are not morally neutral and that it is possible to identify tendencies in them to promote or demote particular moral values and norms" [p. 42]. This means that not merely the usage of a technology has implications for its users, but that the \textit{design} of the technology itself has implications for users and society at large. 

A new and empowering platform therefore requires attention to the role of values. Friedman and Hendry \shortcite{Friedman.2019} propose "value-sensitive design" as a design strategy that considers ethics and moral questions when debating and deciding on a framework for the design of a technology. The point of value-sensitive design is not to dictate a series of acceptable values but rather to provide the means to determine the "right" thing to do in a given context [p. 7]. Following a value-sensitive design approach, we 
propose to center diversity and ethical data collection as core values in our envisioned social media platform. This means that the technology must consider the difference of technology users and connect them with like-minded users or users who complement their skills and interests (section 3.1). 
Simultaneously, this approach brings to focus the privacy and data protection/ownership of users. The data collection takes place inside and outside the EU with the involvement of local experts and ethics committees (section 3.2). A public interest framework and strict data protection regime in Europe mitigate risks of data exploitation and abusive advertising techniques. 

Second, the diversity-aware application (WeNet app) is being designed and developed following a participatory approach. The main idea is to engage with the community of students (as the ultimate beneficiary of the diversity-aware platform) in an iterative process where they are asked to contribute to the design of the application and, in some of the pilot locations, to the analysis of the collected data (see section 3.2). In this way, the students will be made aware of their own data and about the way that it has been used in the application. The ultimate and ambitious goal would be to create a collective identity of the students through their data, an identity that recognizes and welcomes its diverse nature and that could then be conceived as a commons \cite{Morelli.2019}. 




\subsection{3.1 Framework and Operationalization}

In our approach to diversity-aware platform design, we 
see diversity as an opportunity especially for student communities, which are the major end users of the envisioned platform. The idea is that people with diverse backgrounds, experiences, and skills can support each other. At the same time, we acknowledge the difficulty of navigating and coping with increased diversity in our lives. 

Contrary to existing social media platforms, the diversity-aware platform explicitly includes a family of \textit{computational diversity-aware models} supporting human interaction. Learning models construct diversity profiles based on people's past behavior and interactions. A diversity-aware search builds upon these profiles to connect the "right" people together. To support people’s interactions, a diversity alignment mechanism lifts communication barriers to ensure that messages between humans are interpreted correctly, and a diversity-aware incentive mechanism generates incentives to motivate people to support each other. 

To collect the data needed to realize said machine learning algorithms, we need to first define a framework of diversity for the design process and then operationalize the diversity definition. 




\subsubsection{A Framework of Diversity: Empowering the Collective}

A re-design-oriented approach to current social media implies overturning the perspective of fruition: “from a network of computers, which in turn may be connected to people, to a network of people, whose interactions are mediated and empowered by computers”~\cite{divaw}. 
In this perspective, a platform becomes a collectivity enabler, bringing together and supporting the relations of people. Diversity is a prerequisite for this collectivity because a group can only succeed as a community if its members complement each other. 

Collectivity is the set of people which contribute to the achievement of a service, whether they are producers or consumers of the service~\cite{2017-ICCM}. 
For example, in a collectivity like the AirBnB community, the platform enables consumers to find possible places to stay through the people who manage them (producers).
Within a collectivity, not all the members need to know the service to be delivered, but only some of its components. Collectivities have capabilities that are more than the sum of the capabilities of any single member. As a collectivity enabler, the platform acts as a broker in a structural hole~\cite{burt2009structural}, within and among the diversity of collectivities. 

As a broker, the platform favors interactions between people based on a specific need; in other words, the platform is aware of the diversity within and among communities and selects the best suitable collectivity based on a need. As a machine, the platform overcomes the human limits in brokerage. It can reduce the time of search of a need, balance the power in relations, and make available almost unlimited knowledge. Therefore, the platform is a machine that helps humans foster social relations, allowing them to go beyond their physical limitations (e.g., spatio-temporal, computational, memory, interaction). Instead of an obstacle, diversity becomes a lever for the complementarity of people.

\noindent
\subsubsection{Diversity and Diversity-awareness}
Diversity is a complex, multi-layered compositional construct that does not exist \textit{within} individuals. Diversity exists only \textit{between} individuals, namely when two or more individuals interact. This means that we can recognize diversity only when we compare two people and, therefore, when we move at the level of group, organization, community, and society. Moreover, "being different" is a relational concept that applies to everyone, the majority as well as the minority \cite{tsui1991being}.  
While “individual attributes reflect the content of diversity; by contrast, the configuration of attributes within [collectivities] reflects the structure of diversity”~\cite{jackson1995understanding}. This means that diversity attributes can differ in their make-up between various collectivities. In this way, individual differences in attributes create diversity in a collectivity. Figure 1 visualizes this understanding of diversity and its interaction levels.

Diversity-awareness is the ability to cope with this difference across humans, and capitalize on it~\cite{divaw}. 
Diversity-awareness is therefore a human skill that is needed for social interactions. When individuals interact, initial categorizations of the "other" are accompanied by perceptions of similarity or dissimilarity. These perceptions are based on surface-level characteristics (visible attributes like gender, age, etc.) and change when deep-level information (character, personality, skills, abilities) is obtained~\cite{harrison1998beyond}. Over time, as people acquire more information, their perceptions are based more on observed behavior rather than superficial classifications, i.e. stereotypes ~\cite{jackson1995understanding}.


\begin{figure}[t]
\begin{center}

    \begin{minipage}[t]{0.35\textwidth}
        \centering
        \includegraphics[width=\textwidth]{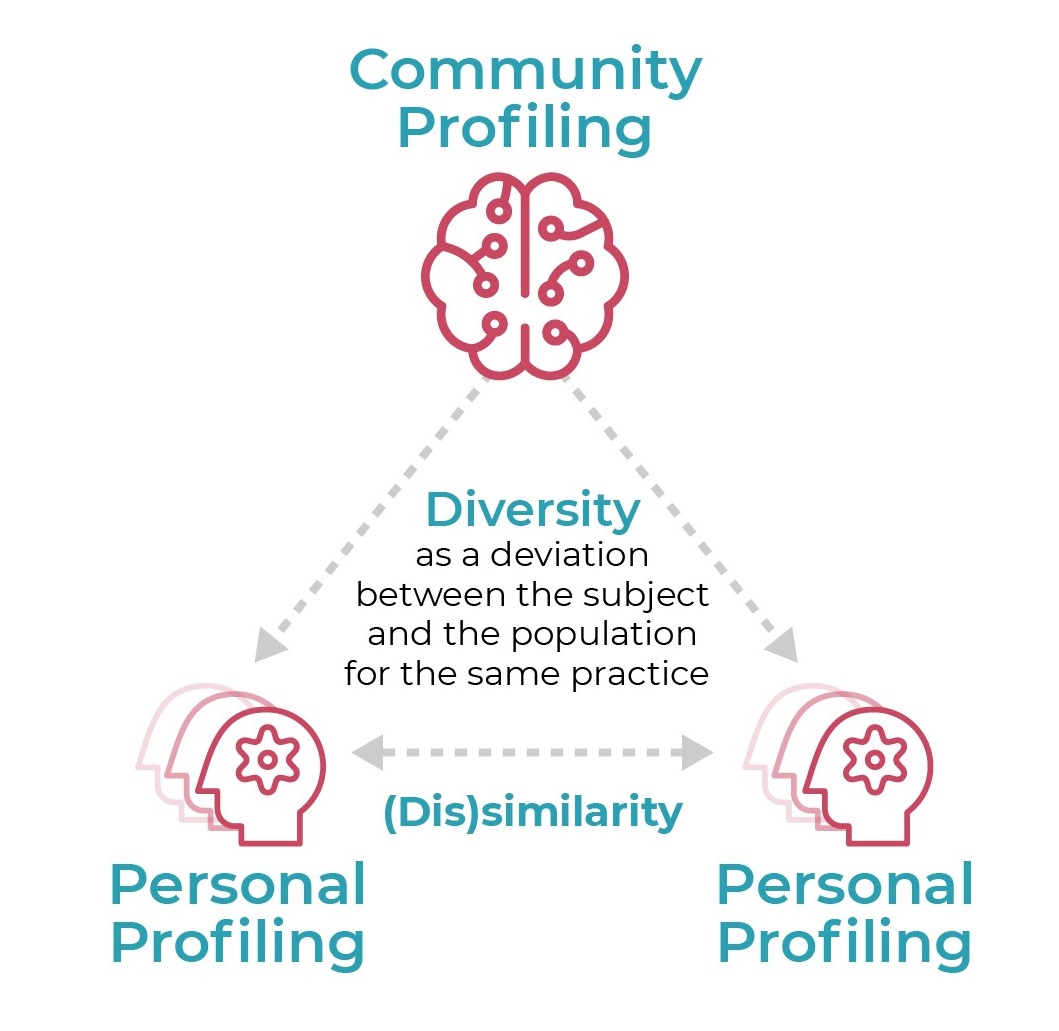}
        \caption{Diversity and interaction levels \cite{D1.1,D1.2}}
    \end{minipage}

\end{center}
\vspace{-0.2 in}
\end{figure}

\subsubsection{Defining and Operationalizing Diversity}

A platform empowering diversity should be understood as a socio-technical system aiming at connecting people for them to achieve their daily life goals. 
In order to do so, the platform must understand people's needs and find a community capable of providing the service that meets the current need. The community then possesses the set of characteristics – a shared practice - that are recognizable on a social level and respond to the expressed need. Let's consider the example of a first-year sociology student who has difficulty with statistics. Since the student needs help in statistics, the platform should choose, among all the communities, the community of people most skilled in statistics. Moreover, the sociology student seeking help should be connected with the most relevant people within the community according to her need (e.g. with someone who exhibits good pedagogical skills or greater skills in sociometry rather than psychometry).

Now, one challenge for the system is to identify the diverse elements of users' social practices.
Alongside the demographic characteristics (here understood as surface diversity), we suggest conceptualizing the diversity of users as "social practices." The theory of social practices~\cite{shove2012dynamics} is proposed as a way to consider both surface level and deep-level characteristics of a person, i.e., to respect both the individual characteristics (e.g. gender, age, etc.) and that of the individual as being part of a collectivity (their skills, abilities, competences).

According to~\cite{Reckwitz.2002}: “a 'practice' … is a routinized type of behavior which consists of several elements, interconnected to one another: forms of bodily activities, forms of mental activities, 'things' and their use, a background knowledge in the form of understanding, know-how, states of emotion and motivational knowledge”. The social practice involves individuals in a specific behavior, that - if repeated - allows the reproduction of the social practice across time and space. So, individuals can be seen as “carriers of practices” who are "recruited" to enact practices~\cite{Reckwitz.2002} according to their background, history, surface and deep-level diversity. A social practice can be further broken down into three fundamental elements \cite{shove2005consumers}: (i) competence, (ii) meaning and (iii) material. 
\begin{itemize}
    \item \textbf{Competence} incorporates skills, know-how, (background) knowledge as well as social and relational skill which are required to perform the practice 
    \item \textbf{Meaning} incorporates the issues relevant to that material, i.e., the understandings, beliefs, value, norms, lifestyle, and emotions
    \item \textbf{Material} covers all physical aspects of the performance of practice, encompassing objects, infrastructures, tools, hardware including the human body
\end{itemize}

\begin{figure*}[tbh]
\begin{center}

    \begin{minipage}[t]{0.7\textwidth}
        \centering
        \includegraphics[width=\textwidth]{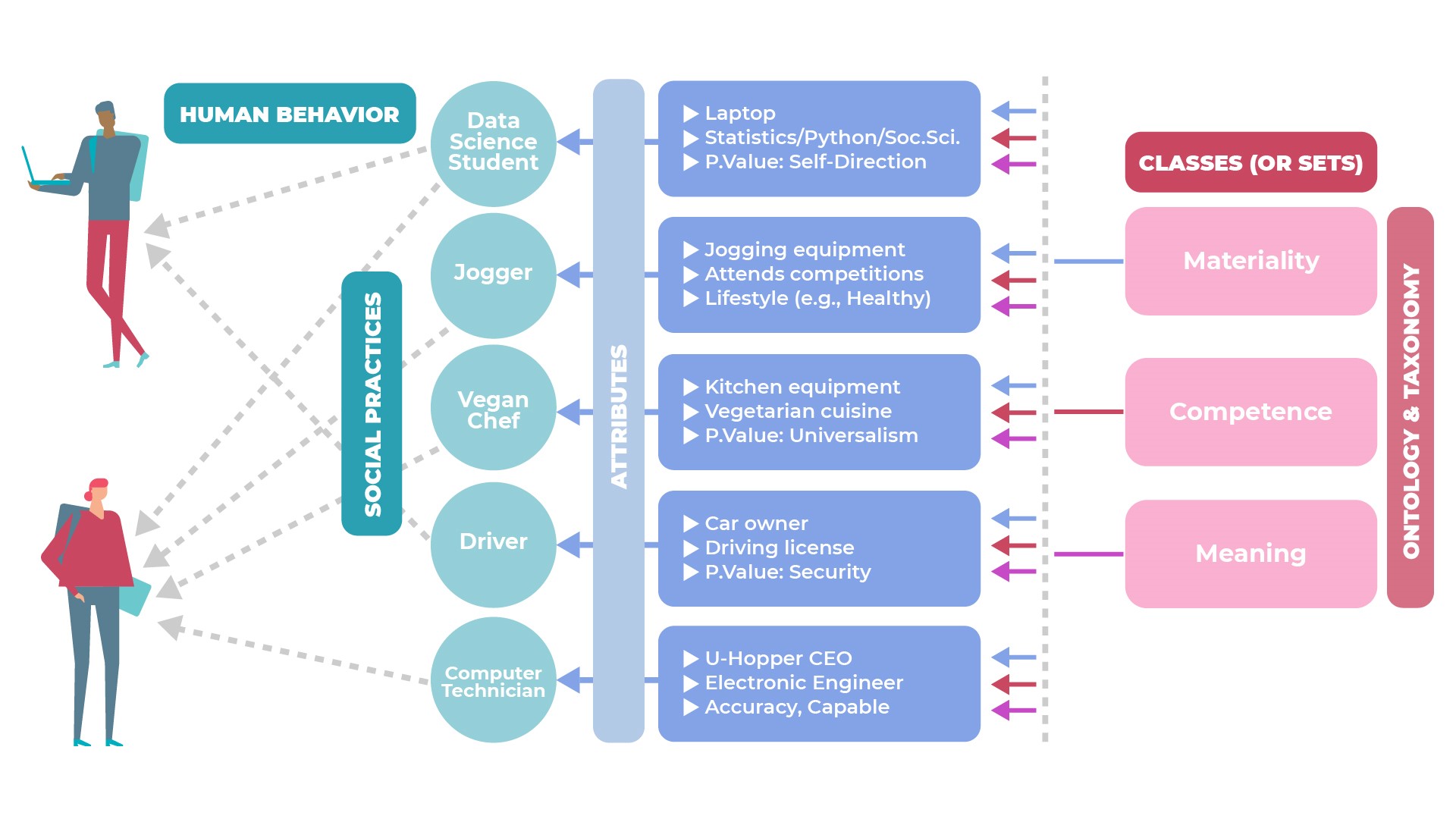}
        \caption{The operationalization of students' diversity as social practices and their components material, meaning, and competence \laura{\textbf{please include citation!} (\textit{citation goes here})}}
    \end{minipage}

\end{center}
\vspace{-0.2 in}
\end{figure*}

Figure 2 visualizes this three-piece concept with the example of student practices. Let us also consider the following example to make the operationalization of diversity as social practices clearer: If we consider the practice of driving “it becomes clear that relations between the vehicle (along with the road and other traffic), the know-how required to keep it in motion and the meaning and significance of driving and passengering are intimately related”~\cite{shove2012dynamics} in forming what~\cite{Reckwitz.2002} calls the 'block' of interconnected elements within but also across these components. 
These three elements exist on a \textit{social} level (that is, separated from the individual). In different combinations, they form the various practices. However, material, competence and meaning can be traced back to the individual. The way an individual combines the elements of a practice reveals their belonging to the practice.

In this sense, individuals are not merely described with skewed attributes, but they are seen as members of a collectivity, also called a community of practice~\cite{wenger1999communities}. They develop a \textit{shared practice}, which becomes a repertoire of resources: experiences, stories, tools, ways of addressing recurring problems. The platform can then help the collectivity of practitioners to improve their performance by leveraging and connecting their different competences, meaning, and material.


\subsection{3.2 Data Collection Process}

In order to develop the machine-learning algorithms that empower users' social interaction and connect users to one another based on their diversity, the project collects its own "diversity" data. This data collection process involves the different universities and student participants recruited to the consortium. It is divided into sets of "pilots," which describe the different location-dependent surveys and testing of the WeNet application \cite{D1.4}. The WeNet application is one of the central products of the research project and connected to the larger social platform under construction. The consortium is diverse by definition, being composed of universities from all over the world: the pilot sites are in Mexico, India, Paraguay, China, Mongolia, England, Italy, and Denmark. A big effort then has been devoted to coordinating and defining the protocols for the data collection. 

Since the pilots are situated in different cultural and socioeconomic contexts, they identify and attend to local specific needs of the student population. This is done by a preparatory exploration of needs and challenges of the respective student population in focus groups and interviews. One example is the Mexico preparatory pilot, where the well-being of students is discussed by reference to food consumption and nutrition. Students in the Mexico preparatory pilot identified eating habits as a challenge in student life, and were interested in contributing to further studies on this topic for the benefit of their own health and their peer community. In that sense, the project regards student participants not merely as data subjects but as active collaborators in the build-up of "their" social platform (participatory design). 

\subsubsection{Pilots: Large-scale Data Collection Activity}

\begin{figure*}[tbh]
\begin{center}

    \begin{minipage}[t]{0.7\textwidth}
        \centering
        \includegraphics[width=\textwidth]{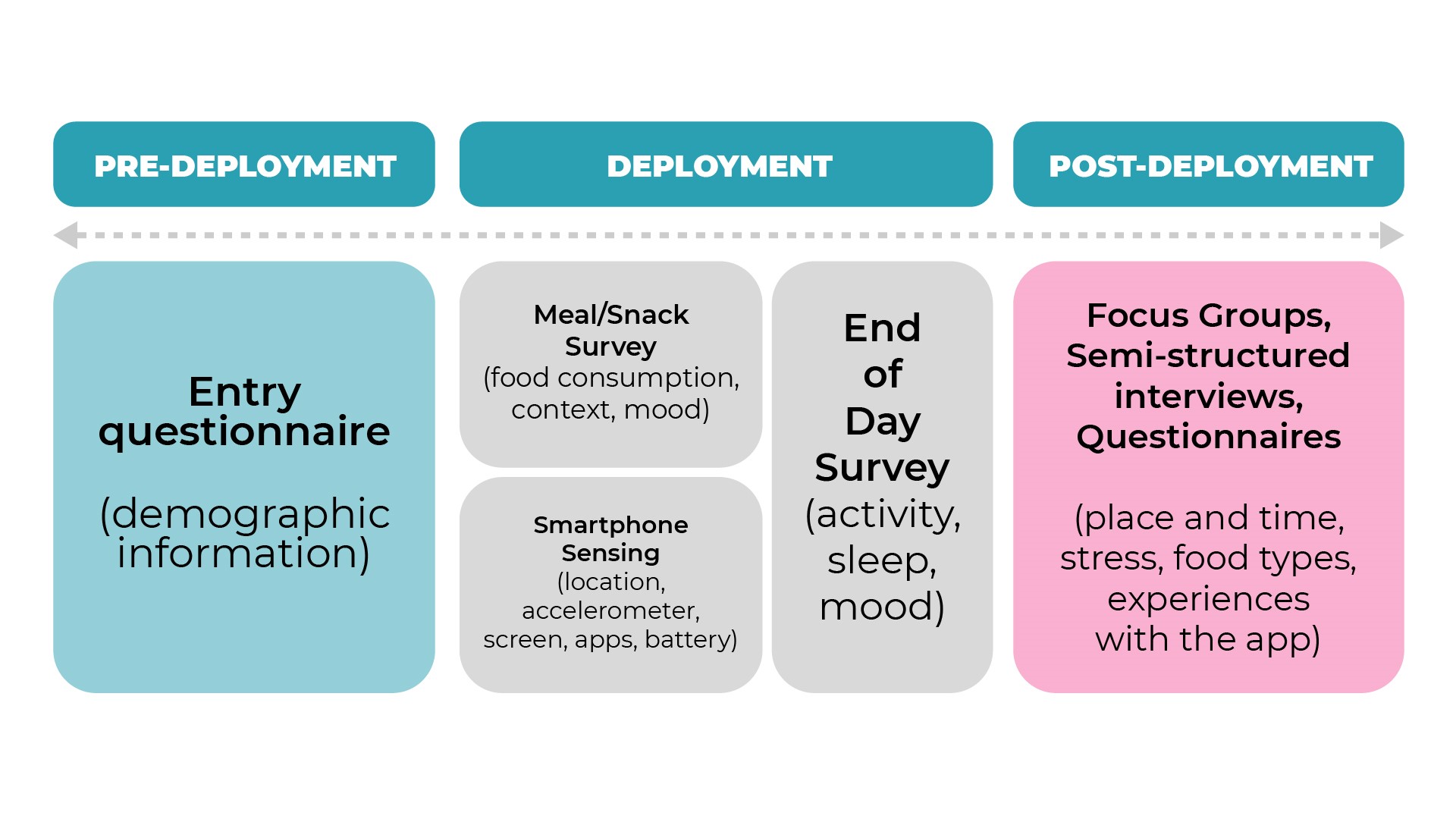}
        \caption{Exploratory data collection in Mexico: mobile app deployment}
    \end{minipage}

\end{center}
\vspace{-0.2 in}
\end{figure*}

Within the project, two different kind of pilots have been designed: the “diversity measurement” pilots, partly done by using the i-Log application \cite{D1.4} 
and the “WeNet application” pilots (where WeNet is built on top of a substantially extended version of the i-Log application). During the four years of the project, 2 diversity measurement pilots and 3 application pilots will be organized. The iterations will allow the researchers to review the implemented model of diversity, improve the machine learning algorithms and the platform integration, and utilize the feedback from the participants to improve the concept. 

The first typology of pilot, the "diversity measurement" pilots, has the aim of collecting the necessary data for the validation of the model of diversity based on social practices and for the training of the algorithms needed to learn those social practices and the students individual and social behaviors. Within the diversity measurement pilots, two main instruments of data collection were used: a survey and i-Log to collect sensor data and text \cite{Zeni.2014}. The survey is distributed online through the LimeSurvey software, while the sensor data is collected via the i-Log application. All the data are anonymized by the partners data controllers and made available to third parties within the project. 

The second typology of pilots, the “WeNet application” pilots, instead aim at testing the diversity-aware algorithms and implementing the model of diversity into an application that mediates the interaction between students.  Within the WeNet application pilots, a chatbot application, designed and developed by the consortium, is used to collect sensor and interaction data. At the end of the experiment (of at least two weeks), a questionnaire is distributed to student participants in order to investigate their experience as users of the application. Following the data minimization principle \cite{Biega.07252020}, only GPS data is collected in order to implement a specific feature in the application. 

All (typologies of) pilots are approved by the local ethical committee of each pilot site, ensuring GDPR compliance. In this respect, the consortium had to face the novelty of the GDPR rules that were challenging universities’ practices. The final goal is to develop a \textit{Research Infrastructure} which will maintain and make available not only the collected data, fully anonymized, but also the documentation of the needed processes which are needed in order to share and (re)use the data collected, thus enabling transparency and replicability. The plan is that the Research infrastructure will make the data available worldwide, also after the end of the project and will also enable more and more projects, like the ones described here, which will then become a source for further research.

\subsubsection{Student Recruitment}

In the pilot activities, students of the different universities are actively involved in qualitative and quantitative data collection. While the "diversity measurement” pilots consider students as the subjects of a scientific investigation into diversity, in the “WeNet application” pilots, students are users of an application that provides specific functionalities. Students then take part in a value co-creation process \cite[p.~13]{Morelli.2020}.

Given the students' active role, their recruitment to the project is essential. Students of course voluntarily participate in the pilots. For the "diversity measurement" pilots, all students regularly enrolled at the different pilot universities are sent an invitation to fill in a web questionnaire implemented in Lime Survey. For very big universities such as the partner university in China, a maximum of 30.000 students were involved. At the end of the main questionnaire, the students can express their interest in participating in the second phase of the data collection, which involves filling in a second questionnaire and using the i-Log application. Among the interested students, for technical reasons, only the ones who declared to have an Android smartphone with an operating system greater than 6.0 are eligible for the subsequent phases of the data collection.

A random sample of 300 students (among those who manifest interest in further cooperation) is selected and sent a request to fill in a second questionnaire and only after completing it, the students receive a password to download and install the i-Log application. In this second phase, it is expected that participants will be paid, if (and only if), at the end of two or four weeks, they have regularly filled in their time diary and kept the i-Log application on. Two weeks after the start of the i-Log data collection, a third questionnaire is sent out in which the participant can express their intention to continue using i-Log for another two weeks.

The students who participate in the above activities and have regularly delivered the data are then invited to participate in the WeNet application pilot. Via the WeNet app, the research team evaluates how students' interaction in the app unfolds while using diversity-aware algorithms to support students’ everyday tasks. A fourth questionnaire at the end of the pilot is sent to users to collect information about the user experience and to collect information to improve usability.

\subsubsection{Exploratory Data Collection: Eating Habits of University Students in Mexico} 


%



Some of the ideas described above were pre-piloted in Mexico. The goal of this initial work was to identify relevant scenarios for the pilot location and to collect initial experimental data. The scenario development was done through interviews and focus groups in order to define the major needs of students in this particular cultural-geographic site \cite{Deliverable7.1} 

\begin{table*}[tbh]
    \centering
    \resizebox{0.65\textwidth}{!}{%
    \begin{tabular}{l c c c l} 
    
    \hline 
    
    \rowcolor{Gray}
    \textbf{Phase} & 
    \textbf{\# of Days} &
    \textbf{\# of Workshop Participants} & 
    \textbf{\# of Recruited Volunteers} &
    \textbf{Student Population}
    \\ [0.5ex] 
    \hline
     
    I &
    37 &
    32 &
    29  &
    Universidad Autonoma de San Luis Potosi \\ 

    \rowcolor{gray!5}
    II &
    23 & 
    90 &
    55 &
    Universidad Tangamanga \\

    Total &
    60 & 
    122 &
    84  &
      \\
     
    \hline 
    \end{tabular}
    }
    \caption{Summary of participants in Mexico study}
    \vspace{-0.2 in}
\end{table*}

A clear focus emerged in the Mexico pilot, where students face particular challenges when it comes to healthy eating. The subsequent collection of data was done via the i-Log application.
In Mexico, the number of people affected by overweight and obesity has increased dramatically to the point of being officially declared an epidemiological emergency \cite{Gob2016}. Recent national data shows that overweight and obesity affect over 70\% of adults \cite{Gob2016}. 
Childhood obesity is a similar challenge, placing Mexico among the countries with highest prevalence of childhood obesity worldwide \cite{UnicefMexico2020}. Obesity and overeating  are major risk factors for a number of chronic diseases, including diabetes, cardiovascular diseases, and cancer, and  also contribute to low self-esteem and depression \cite{Bray1992,Stunkard2003}. 
Since overweight and obesity are critical challenges faced by Mexican young adults \cite{Mohan2020,Turnbull2019}, we expected university students to be motivated by this topic and open to reflection about their eating and physical activity habits \cite{Richards2006,Robinson2013}. 

The preparatory pilot study included participants from San Luis Potosi City in Mexico (ca. 1.2 million inhabitants). The city is home to several universities. Together with our local collaborators, we recruited students from two universities  \cite{Meegahapola.2021.b}. The recruitment campaign was launched in June 2019. First, we conducted two workshops in August and October 2019, in which study goals, data collection and processing procedures, and data and privacy protection protocols were explained to attendees. 
We used a basic screening criteria to select participants for the study, including ownership of an android mobile phone, and not having eating disorders like anorexia or bulimia. After the screening process, participants filled in a consent form and a pre-deployment questionnaire to capture basic demographic details. Afterwards, participants installed the i-Log mobile app in their phones, which would send notifications to participants requesting them to report their meals and snacks and the corresponding context throughout the day, while recording sensor data.

The first phase ran in September-October 2019. A second phase ran in November-December 2019. In total, data were collected over 60 days and involved 84 participants. The types of collected data types are summarized in Figure 3. The participation summary during the two phases is described in Table 1. Overall, out of a total of 7898 sent notifications, the response rate by participants was 49.3\%. Filtering out incomplete reports resulted in a total of 3278 complete food consumption reports, which included 1911 meals and 1367 snacks. At the end of the mobile deployment phase, an eight-person focus group and ten semi-structured interviews were conducted, and a short questionnaire with open-ended questions was administered to a subset of participants. The goals were to understand  possible links between food consumption and other variables (e.g., place, time, or mood); and the user experience and perceived value of the app.

Results from the initial data analysis showed that  smartphone sensing can be used to infer self-perceived levels of eating behavior with an accuracy of 87\% in a three-class inference task \cite{Meegahapola.2021.b}. In addition, sensing can be used to analyze the social contexts in which students eat during the day, and infer a basic classification of social eating (eating alone or with others) with an accuracy of 84\% \cite{Meegahapola2020b}. 
Some of these initial results could be used as part of community reflections, where participants could elaborate on the possible benefits that learning about community patterns (e.g. the practice of social eating) could bring to their own reflective process. This issue has key relevance for our ongoing work.

\section{4. Ethical Challenges}

Whereas we consider diversity-aware platform design an innovation vis à vis existing social media designs, there are still ethical challenges (see figure 4). These ethical challenges require particular attention in the design process to avoid repeating old mistakes or making new ones.

\subsection{Diversity and Bias}
Leveraging diversity in a technology means quantifying and "reducing" diversity to measurable or even numeric categories. This task - beyond being demanding - risks introducing static and flat categories that ironically fail to capture the diversity of people. For instance, transgender people refuse to be captured by any label. The diversity of gender identities then relies on some individual \textit{not} being quantifiable or countable ~\cite{Keyes.2019}. On an ontological level, seizing the diversity of transgender individuals by documenting them as label "x" or label "y" is a violent act that prevents them from \textit{being} themselves or embracing their self/their identity. An important question for diversity-aware platform design is then: how can we define and leverage the diversity of users in a way that does not introduce discrimination and bias?

One solution is to define diversity in a way that captures social characteristics and practices that are transversal, or general, to people, without these being generic or discriminating. In our case, this means finding the right level of  \textit{granularity} of students' social practices. On the one hand, defining a social practice too broadly (e.g. "walking") risks including everyone, without the possibility to represent diversity. On the other hand, defining a practice too specifically (e.g. walking with shoes on) incurs potential discrimination, as some individuals may enjoy e.g. barefoot running.

This brings us to another ethical concern about the operationalization of diversity: the ability of a system to capture "truly diverse" practices that deviate from the norm.
Social practices usually constitute the dominant behavior in a society. They are considered "normal" routines, which "are typically and habitually performed in (a considerable part of) a society" \cite{Holtz.2013}. When a collectivity engages in diverse social practices, the risk is that the dominant part of the collectivity "dictates" what is perceived as common practices, or simply as practices. Minority practices that deviate from the norm may not be captured in the data collection or promoted by the algorithm ("threat of invisibility" \cite{Bucher.2012}, "statistical stereotyping” \cite[p. 171]{CheneyLippold.2011}, cf. \cite{Schelenz.2019}). 
This concern is amplified with regard to machine learning algorithms because they are very good at recognizing patterns but less well equipped to capture nuances ~\cite{Matzner.2019}. A related concern is that the system may not be able to detect new and emerging social practices that have no documentation in historical data.

Furthermore, social practices do not originate in a vacuum, they are outcomes of societal interactions and carry norms, values, and expectations. They also carry gender and racial biases. For example, when we think of the sport of ballet or dancing, we usually envision girls (and probably White girls) enacting the practice. When we think of soccer, we usually think of boys or men. The practice of rapping is more associated with Black people than White people, and again Black males rather than Black females. Here, we can see that potentially neutral and descriptive diversity categories are not so neutral \cite{Haslanger.2012}. 
While the computer system merely reflects societal biases (in other words, these biases do not \textit{originate} in the platform), it is important to ask whether a diversity-aware platform must mitigate pre-existing biases via machine learning fairness ~\cite{Oneto.2020} or other measures. This is a matter for future research.



\subsection{The Limits of Diversity}

Despite the rhetorical "celebration" of diversity, diversity can also be problematic in some instances. This relates to 1) harmful social practices and 2) harmful interactions between users in the platform. On the one hand, the system may promote socially unacceptable or dangerous practices because users in the platform exhibit them routinely. Think of binge drinking, for example. Excessive alcohol consumption among students has become a public health problem but is so widespread that it amounts to "an organising principle of university social life" ~\cite{Supski.2017}. Is it ethical for the system to promote this diversity aspect in a student community? Here, we see that promoting diversity is not a neutral task but requires balancing values in a collectivity. The system should be transparent about the limits of diversity that designers or the collectivity itself choose to impose. 

On the other hand, there can be hostility between individuals. 
In some instance, users may not tolerate the kind of diversity that the system enables. In extreme cases, users may even misuse the platform to spread hate and division. Therefore, it is important that diversity-aware design sets boundaries regarding the rules of behavior. While free speech is a protected right, hate speech is a crime that unfortunately lacks regulation and prosecution in existing social media platforms \cite{Lumsden.2019,Daniels.2009}. Diversity-aware platforms must attend to hate speech and sexist or racist violence in the platform because it otherwise constitutes a structural barrier for different groups to interact online. The diversity-aware platform may even consider creating “safe spaces" for some users.

\subsection{Diversity-Aware Data Collection}

The data collection process presented in this paper is situated in a machine learning context where technical views of diversity have been proposed \cite{Gong2019v2}. New approaches to diversity-aware machine learning focus on data diversification, model diversification, and inference diversification. Diversity-aware machine learning builds on the availability of data. This offers new opportunities to learn about the diversity of users, e.g. diverse aspects of students' well-being \cite{Meegahapola.2021}. 
However, there is a set of limitations and open questions that require future work.

First, data collected for a diversity-aware platform must reflect the diversity of a population (e.g. in our case, the diversity of university students.) Two issues raise concerns. If a data set represents a population of diverse people but each sub-group is represented by the data of only a few individuals, models may learn and perform poorly, leading to biased or inaccurate outcomes. In other words, a few individuals of a group cannot represent the entire group. There is diversity within groups and sub-groups. It is therefore important to ensure a representative and large enough sample to expect to capture a wider view of the diversity of a population. Another problem occurs when the data set is not diverse enough by design. For example, social practices of university students may differ based on their disciplines. If we then only recruit students from the psychology department for an experiment, the resulting data may only reflect the diversity of students in psychology. In this case, it is not necessarily valid to draw conclusions about other students at the same university, i.e., about the entire collective.

Second, data that informs the machine learning algorithms of a diversity-aware platform should reduce gender bias. Gender dynamics and historical forms of gender discrimination may influence the experiences and well-being of students. Gender data bias has been identified as a barrier to fair treatment and outcomes from computer systems \cite{CriadoPerez.2019,Buolamwini.2018}.  
It is thus crucial to build models on data sets that represent gender diversity.

\begin{figure}[t]
\begin{center}

    \begin{minipage}[t]{0.3\textwidth}
        \centering
        \includegraphics[width=\textwidth]{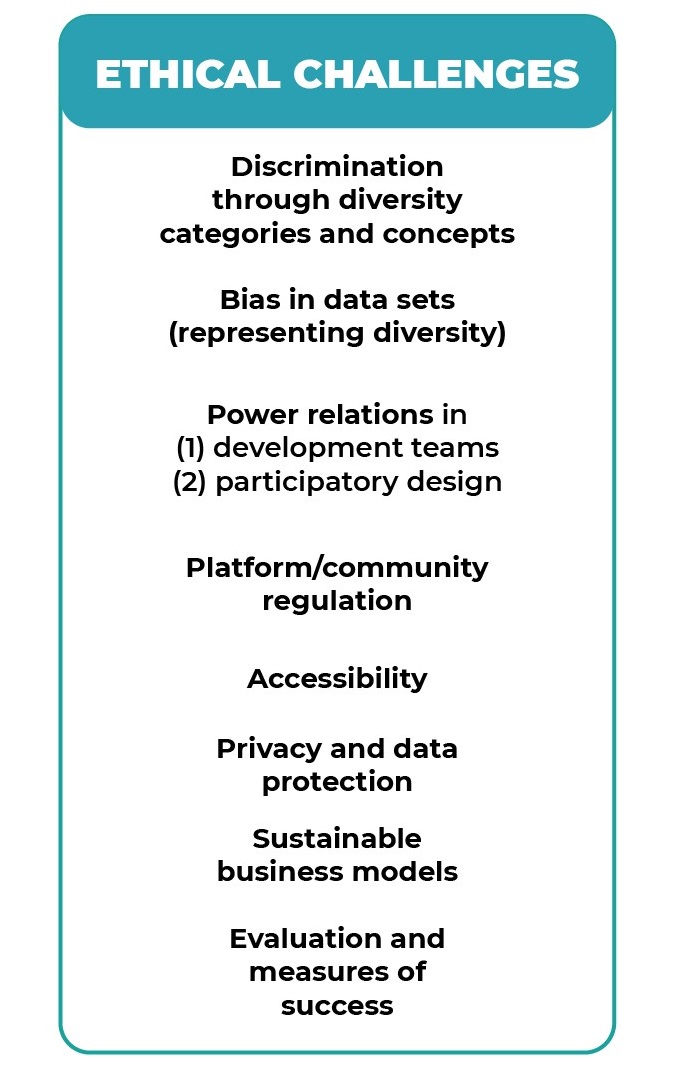}
        \caption{Ethical challenges of diversity-aware platform design}
    \end{minipage}

\end{center}
\vspace{-0.2 in}
\end{figure}

Third, diversity-aware data collection must take into account local habits, which may differ across geographical and cultural regions. Collecting smartphone data requires insights into local practices, as the use of smartphones among young people differ across countries \cite{Mathur2017,Meegahapola.2021}. Students may use distinct apps in some countries, or use mobile services differently because of the cost of phone devices and data plans, popular local trends, and culture at large \cite{Meegahapola2020a,Meegahapola2020b}. Data collection needs to take these differences across cultures and countries into account.




Finally, smartphone data sets in academic research remain small compared to data sets in other domains in computing, like computer vision or natural language processing. We believe that smartphone data collection efforts across multiple geographical regions with diverse populations, as we present here, might motivate the further development of diversity-aware machine learning and inference techniques, and contribute to the field while raising awareness about other potential issues with data sets. 

\subsection{Participatory Approaches and Diversity}
Applying a  participatory approach to data collection and analysis can ensure that the diversity that is detected by the algorithms is the one that better represent the community of students. The students themselves, who are active participants rather than passive data subjects, can help define their diversity through data and suggest how to use it in diversity-aware services. Participatory design has in general become a prominent design strategy for human-centered technology \cite{Simonsen.2013}, for instance in living labs for co-design and co-creation \cite{Baelden.2015}. 

Such an approach though requires a long term engagement with the users/students that is difficult to achieve within the limited frame of  a research project. Moreover, power relations and structural barriers may influence students' ability to participate in the design process. Work load, stress and mental health problems, but also care work responsibilities (students may take care of children or elderly people) prevent students from being able to dedicate time and energy to their participation. Especially students who are structurally disadvantaged in society (single parents, immigrants, students with disabilities) may lack the assistance they need to realize their participation despite the \textit{formal} opportunity to participate. Diversity-aware design processes should thus attend to these issues during the recruitment of student participants when they use participatory design approaches. 

\subsection{Business Models for Diversity-aware Platforms}
In existing platforms, the pressure to monetize the service, e.g. via advertisement, has been identified as a major driver of suppressing diversity in a platform ~\cite{Zuboff.2019,Susser.2019}, see section 2. However, a platform cannot exist from the mere \textit{aspiration} to connect users and improve their well-being. Beyond public funding, a sustainable diversity-aware platform must then consider an economic framework that does not compromise efforts to promote diversity and data protection. One solution may be to empower the online community to manage their own data as a commons ~\cite{Ostrom,Morelli.2019}.

\subsection{Accessibility}
A diversity-aware platform for social relations must also consider the accessibility of the service to users with disabilities. In that sense, diversity is not just a conceptual tool or fact of life that the technology leverages. Diversity is an intrinsic and instrumental value which demands that we accommodate users with different needs \cite{Schelenz.2019}. For users with disabilities, this means that the platform must remove barriers for them to consume the service. For instance, blind users depend on assistive technologies such as a screen reader. The diversity-aware platform must then be compatible with the technical standards to support the use of a screen reader \cite{Hamdy.2020}.

\subsection{Measuring the success of the platform}
Finally, an ethically inspired diversity-aware platform is expected to answer questions that researchers and users have for existing platforms. But how can we measure the advantages of a diversity-aware paradigm vs. previous design approaches? Certainly, the diversity-aware algorithms are expected to perform better than traditional models. Hence, user experience and satisfaction must increase vis à vis existing solutions. However, a diversity-aware platform provides more than a service. It is a place for community, belonging, empowerment (in terms of increased autonomy), and personal growth. Some of the benefits may be immaterial and difficult to measure. What is the threshold for success of a diversity-aware platform and who defines success? These are open questions for future elaborations. 



\section{5. Conclusion}
This paper has presented a diversity-aware paradigm for the design of social media platforms. Considering the cluster of critique surrounding existing platforms, there is momentum for a change in perspective. The design approach offered in this paper leverages the diversity of users to their benefit, which we label "diversity-aware platform design." This design approach is at the heart of a European technology development project (anonymized for peer review) that seeks to design a platform and corresponding app which fosters students' social relations and well-being. 

We have laid out the theory and practice of 1) framing and operationalizing "diversity", and 2) collecting "diversity" data in compliance with strict data protection regulations, GDPR \cite{EuropeanParliament.2016}. In order to define the diversity of a community, we propose to look at the social practices present in said community. These practices can be learned from "diversity" data, which our team collected from surveys and the interaction of students with an application. Lastly, we have discussed the ethical challenges of diversity-aware platform design. Although diversity invokes a sense of inclusion and fairness, a diversity-aware platform is not \textit{automatically} free from bias. The design community then has to constantly reflect on the implications of designing for diversity.

\section{Acknowledgments}
This research has received funding from the EU Horizon 2020 project “WeNet – The Internet of Us” under grant agreement No 823783. The authors have contributed equally to the paper under the coordination of the ethics researcher. The research presented here reflects the work of a much larger consortium. We would especially like to acknowledge the contributions of Luca Cernuzzi, Shyam Diwakar, Kobi Gal, Amarsanaa Ganbold, George Gaskell, Jessica Heesen, Loizos Michael, Daniele Miorandi, Carles Sierra, and Donglei Song. 


\bibliographystyle{aaai} \bibliography{references}


\end{document}